\documentclass[english,english,reprint,jcp,aip,superscriptaddress]{revtex4-1}
\usepackage[T1]{fontenc}
\usepackage[latin9]{inputenc}
\usepackage{array}
\usepackage{multirow}
\usepackage{tipa}
\usepackage{tipx}
\usepackage{amsmath}
\usepackage{graphicx}
\usepackage{esint}

\makeatletter

\providecommand{\tabularnewline}{\\}

\@ifundefined{textcolor}{}
{%
 \definecolor{BLACK}{gray}{0}
 \definecolor{WHITE}{gray}{1}
 \definecolor{RED}{rgb}{1,0,0}
 \definecolor{GREEN}{rgb}{0,1,0}
 \definecolor{BLUE}{rgb}{0,0,1}
 \definecolor{CYAN}{cmyk}{1,0,0,0}
 \definecolor{MAGENTA}{cmyk}{0,1,0,0}
 \definecolor{YELLOW}{cmyk}{0,0,1,0}
}

\usepackage{babel}

\usepackage{babel}

\usepackage{babel}

\makeatother

\usepackage{babel}
\begin{document}

\title{Deviations from piecewise linearity in the solid-state limit with
approximate density functionals}

\author{Vojt\v{e}ch Vl\v{c}ek}

\homepage{These two authors contributed equally to this work.}

\affiliation{Bayerisches Geoinstitut, Universität Bayreuth, D-95440 Bayreuth,
Germany}

\affiliation{Fritz Haber Center for Molecular Dynamics, Institute of Chemistry,
The Hebrew University of Jerusalem, Jerusalem 91904, Israel}

\author{Helen R. Eisenberg }

\homepage{These two authors contributed equally to this work.}

\affiliation{Fritz Haber Center for Molecular Dynamics, Institute of Chemistry,
The Hebrew University of Jerusalem, Jerusalem 91904, Israel}

\author{Gerd Steinle-Neumann}

\email{g.steinle-neumann@uni-bayreuth.de}

\affiliation{Bayerisches Geoinstitut, Universität Bayreuth, D-95440 Bayreuth,
Germany}

\author{Leeor Kronik}

\email{leeor.kronik@weizmann.ac.il}

\affiliation{Department of Materials and Interfaces, Weizmann Institute of Science,
Rehovoth 76100, Israel}

\author{Roi Baer}

\email{roi.baer@huji.ac.il}

\affiliation{Fritz Haber Center for Molecular Dynamics, Institute of Chemistry,
The Hebrew University of Jerusalem, Jerusalem 91904, Israel}
\begin{abstract}
In exact density functional theory (DFT) the total ground-state energy
is a series of linear segments between integer electron points, a
condition known as ``piecewise linearity''. Deviation from this
condition is indicative of poor predictive capabilities for electronic
structure, in particular of ionization energies, fundamental gaps,
and charge transfer. In this article, we take a new look at the deviation
from linearity (i.e., curvature) in the solid-state limit by considering
two different ways of approaching it: a large finite system of increasing
size and a crystal represented by an increasingly large reference
cell with periodic boundary conditions. We show that the curvature
approaches vanishing values in both limits, even for functionals which
yield poor predictions of electronic structure, and therefore can
not be used as a diagnostic or constructive tool in solids. We find
that the approach towards zero curvature is different in each of the
two limits, owing to the presence of a compensating background charge
in the periodic case. Based on these findings, we present a new criterion
for functional construction and evaluation, derived from the size-dependence
of the curvature, along with a practical method for evaluating this
criterion. For large finite systems we further show that the curvature
is dominated by the self-interaction of the highest occupied eigenstate.
These findings are illustrated by computational studies of various
solids, semiconductor nanocrystals, and long alkane chains. 
\end{abstract}
\maketitle

\section{Introduction}

Kohn-Sham (KS) density functional theory (DFT) \cite{Hohenberg1964,Kohn1965}
is a widely used first-principles approach to the many-electron problem.
It is based on mapping the system of $N$ interacting electrons into
a unique non-interacting system with the same ground state electron
density.~\citep{Levy1982,Lieb1983} In the non-interacting system
the density is determined by $n\left(\mathbf{r}\right)=\sum_{i}f_{i}\left|\psi_{i}\left(\mathbf{r}\right)\right|^{2}$
where $\psi_{i}\left(\mathbf{r}\right)$ $\left(i=1,2,...\right)$
are normalized single particle eigenstates and $f_{i}$ are the corresponding
occupation numbers. The eigenstates are determined from the KS equations
\begin{equation}
\hat{H}\psi_{i}=\varepsilon_{i}\psi_{i},\label{eq:KS Eqs}
\end{equation}
where $\varepsilon_{i}$ are the (monotonically increasing) KS eigenvalues
(see footnote %
\footnote{For spin unpolarized (polarized) KS systems the value of the occupation
number $f_{i}$ is equal to $2\,\left(1\right)$ if $\varepsilon_{i}<\varepsilon_{H}$,
equal to $0$ if $\varepsilon_{i}>\varepsilon_{H}$ and $0\le f_{i}\le2\left(1\right)$
if $\varepsilon_{i}=\varepsilon_{H}$, where $\varepsilon_{H}$, the
highest occupied eigenvalue, is determined such that $\sum_{i}f_{i}$
is equal to the total number of electrons $N=\int n\left(\mathbf{r}\right)d^{3}r$.
The lowest eigenvalue for which $f_{i}=0$ is referred to as $\varepsilon_{L}$. %
}) and 
\begin{equation}
\hat{H}=-\frac{1}{2}\nabla^{2}+v_{H}\left({\bf r}\right)+v_{XC}\left({\bf r}\right)+v_{ext}\left({\bf r}\right)\label{eq:KS H}
\end{equation}
is the KS Hamiltonian (atomic units are used throughout). In Eq. \eqref{eq:KS H},
$v_{H}\left({\bf r}\right)$ is the Hartree potential, $v_{XC}\left({\bf r}\right)$
the exchange-correlation (XC) potential and $v_{ext}\left({\bf r}\right)$
is the external potential operating on the electrons in the interacting
system. While DFT in general, and the KS equation in particular, are
exact in principle, the XC potential functional is always approximated
in practice and thus defines the level of theory applied.

The exact XC energy functional, $E_{XC}\left[n\right]$, from which
the XC potential is derived via the relation $v_{XC}\left({\bf r}\right)=\delta E_{XC}\left[n\right]/\delta n\left({\bf r}\right)$,
is known to satisfy a number of constraints (e.g., Ref.~\citenum{PerdewKurthBookChap}).
One constraint, on which we focus here, is the \emph{piecewise-linearity}
property.~\citep{Perdew1982a} Perdew \emph{et al.} have argued that
the ensemble ground-state energy \emph{$E\left(N\right)$ }as a function
of electron number, $N$ where $N_{0}-1\le N\le N_{0}$, must be a
series of linear segments between the integer electron points $N_{0}$.
Within the KS formalism this requirement translates directly into
a condition on the XC energy functional, $E_{XC}\left[n\right]$.

An important manifestation of piecewise-linearity is the relation
between the highest occupied eigenvalue, $\varepsilon_{H}$, and the
ionization potential, $I(N_{0})\equiv E\left(N_{0}\right)-E\left(N_{0}-1\right)$.
These considerations have been originally developed for finite systems;
infinite systems are discussed in detail below. For the exact functional,
piecewise-linearity dictates that $I=-dE/dN$. In addition, Janak's
theorem \citep{Janak1978} states that for any (exact or approximate)
XC functional, the highest occupied eigenvalue obeys 
\begin{equation}
\varepsilon_{H}=\frac{dE_{KS}}{df_{H}},\label{eq:Janak}
\end{equation}
where $E_{KS}$ is the KS estimate for the energy of the interacting
system. For any change in electron number $N$, the same change occurs
in $f_{H}$, the occupation number of the highest occupied eigenstate
of the non-interacting system. Thus we find the result $I=-\varepsilon_{H}$
for a KS theory which uses the exact XC functional (i.e. for which
$E_{KS}=E$). This exact condition, known as the \emph{ionization
potential theorem},~\citep{Perdew1982a,Levy1984a,Almbladh1985,Perdew1997}
can be conveniently restated in terms of the energy curvature, $C$,
defined as the second derivative of the total energy functional with
respect to the fractional electron number, 
\begin{equation}
C=\frac{d^{2}E}{dN^{2}}=\frac{d^{2}E_{KS}}{df_{H}^{2}}=\frac{d\varepsilon_{H}}{df_{H}},\label{eq:Curvature Defnition}
\end{equation}
where Janak's theorem has been used in the third equality. Fulfillment
of piecewise-linearity implies that $C=0$, i.e. that the curvature
is zero.

Despite the importance of piecewise-linearity, it has long been known
that standard application of commonly used functional classes, such
as the local density approximation (LDA), the generalized gradient
approximation (GGA), or conventional hybrid functionals with a fixed
fraction of Fock exchange (e.g. Ref.~\citenum{Becke1993a}), grossly
disobeys this condition. In practice, a substantial, non-zero curvature
is observed. The $E_{KS}\left(f_{H}\right)$ curve is typically strongly
convex (see, e.g.~\citenum{Mori-Sanchez2006,Ruzsinszky2007,Vydrov2007,Mori-Sanchez2008a,Cohen2008c,Cohen2008a,Haunschild2010many,srebro2012does,gledhill2013assessment,Cohen2014Dramatic})
and, correspondingly, $-\varepsilon_{H}$ can underestimate $I$ by
as much as a factor of two.~\citep{Tozer1998,Allen2002}

The lack of piecewise-linearity in approximate functionals further
affects the prediction of the fundamental gap, $E_{g}$, defined as
the difference between the minimum energy needed for electron removal
and the maximum energy gained by electron addition. Even with the
exact functional, the KS eigenvalue gap, $\varepsilon_{L}-\varepsilon_{H}$
(where $\varepsilon_{L}$ is the energy of the lowest unoccupied eigenstate),
need not equal $E_{g}$ \citep{Perdew1983,Sham1983}. Instead, 
\begin{equation}
E_{g}=\varepsilon_{L}-\varepsilon_{H}+\Delta_{XC},\label{eq:Gaps_relations_exact}
\end{equation}
where $\Delta_{XC}$ is the derivative discontinuity~\citep{Perdew1982a,Perdew1985,Sagvolden2008,Cohen2014Dramatic}
- a spatially-constant ``jump'' in the XC potential as the integer
number of particles is crossed. This discontinuity is itself a consequence
of piecewise linearity: The discontinuous change of slope in the energy
as a function of electron number must also be reflected in the energy
computed from the KS system. Some of it is contained in the kinetic
energy of the non-interacting electrons, but the rest must come from
a discontinuity in the XC potential.~\citep{Perdew1982a} Note that
within the generalized KS (GKS) scheme (see footnote %
\footnote{where the interacting-electron system is mapped into a partially interacting
electron gas that is still represented by a single Slater determinant~\citep{Seidl1996}.%
}) part of the discontinuity in the energy may also arise from a non-multiplicative
(e.g., Fock) operator.~\citep{Seidl1996,KronikKuemmel2008,Yang2012}
Therefore the derivative discontinuity in the XC potential may be
mitigated and in some cases even eliminated.~\cite{Eisenberg2009,Stein2010,Kronik2012,Yang2012}

For any approximate (G)KS scheme, $E_{g}$ can be expressed as \citep{Stein2012}
\begin{equation}
E_{g}=\varepsilon_{L}-\varepsilon_{H}+\frac{1}{2}\left(C^{hole}+C^{elec}\right)+\Delta_{XC},\label{eq:Gaps_relations}
\end{equation}
where $C^{hole}$ and $C^{elec}$ are the curvatures associated with
electron removal and addition, respectively. The curvatures act as
``doppelgänger'' for the missing derivative discontinuity. Whereas
in the exact functional all curvatures are zero and the difference
between $E_{g}$ and the eigenvalue gap is given solely by $\Delta_{XC}$,
for standard approximate (semi-)local (LDA and GGA) or hybrid functionals,
employed in the absence of ensemble corrections, $\Delta_{XC}$ is
zero and the addition of the average curvature compensates quantitatively
for the missing derivative discontinuity term.~\citep{Stein2012}
In the most general case, both a remaining curvature and a remaining
derivative discontinuity will contribute to the difference between
$E_{g}$ and $\varepsilon_{L}-\varepsilon_{H}$.

For small finite systems, the criterion of piecewise linearity (i.e.,
zero curvature) has been employed to markedly improve the connection
between eigenvalues and ionization potentials or fundamental gaps,
and often also additional properties, in at least four distinct ways:
(i) In the imposition of various corrections on existing underlying
exchange-correlation functionals;~\citep{cococcioni2005linear,LanyZunger2009,Dabo2010,Stein2012,
ChaiChen2013,Ferretti2014}(ii) In first-principles ensemble generalization of existing functional
forms;~\citep{kraisler2013piecewise, kraisler2014fundamental} (iii)
In the construction and evaluation of novel exchange-correlation functionals;~\citep{Kuisma2010,ArmientoKuemmel2013}
And (iv) in non-empirical tuning of parameters within hybrid functionals,~\citep{Sai2011,Atalla2013}
especially range-separated ones.~\citep{Stein2010,Baer2010a,Kronik2012,Srebro2014}

Unfortunately, this remarkable success of the piecewise-linearity
criterion does not easily transfer to large systems possessing delocalized
orbitals. For example, for a LDA treatment of hydrogen-passivated
silicon nanocrystals (NCs), the fundamental gap computed from total
energy differences approaches the KS eigenvalue gap with increasing
NC size.~\citep{Ogut1997,Godby1998} As mentioned above, for LDA
$\Delta_{XC}=0$. Taken together with Eq.~\eqref{eq:Gaps_relations},
this implies that as system size grows the average curvature becomes
vanishingly small and piecewise linearity is approached.~\citep{Mori-Sanchez2008a}
Despite this, the ionization potential obtained this way does not
agree with experiment.~\citep{salzner2010modeling}

This limitation is intimately related to the vanishing ensemble correction
to the band gap of periodic solids~\citep{kraisler2014fundamental}
and even to the failure of time-dependent DFT for extended systems
\citep{Onida2002,IzmaylovScuseria2008}. This is a disappointing state
of affairs, because the zero curvature condition that has been used
so successfully for small finite systems, both diagnostically and
constructively, appears to be of little value for extended systems,
even though the problem it is supposed to diagnose is still there.

In this article, we take a fresh look at this problem, by considering
the evolution of curvature with system size. We approach the bulk
limit in two different ways: (i) Calculations for an increasingly
large but finite system (namely nanocrystals and molecular chains).
(ii) Calculations for a crystal represented by an increasingly large
reference cell with periodic boundary conditions. We show that in
both cases the curvature approaches zero. However, it doesn't do so
in same fashion, due to the presence of a compensating background
charge in the periodic system. Based on these findings, we present
a new criterion for functional construction and an assessment derived
from the size-dependence of the curvature, along with a practical
method for evaluation this criterion. We further show that the curvature
for large finite systems is dominated by the self-interaction of the
highest occupied eigenstate. These findings are illustrated by computational
studies of semiconductor NCs and long alkane chains.

\section{\label{sec:EC-finite}Energy curvature in large finite systems}

\subsection{\label{sub:FinSys-General-considerations}General considerations}

We first examine finite systems, in which, as noted above, curvature
effects have been already studied extensively. As a first step in
our general theoretical considerations, we express the curvature of
a finite system as the rate of change in the energy of the highest-occupied
KS-eigenstate as an electronic charge $q$ is removed or added (see
footnote %
\footnote{In this paper all systems are treated strictly in the closed shell
spin-unpolarized ensemble, so any removal or addition of small amounts
of electronic charge preserves the unpolarized spin nature of the
system. %
}) to the system: 
\begin{equation}
C=\frac{d\varepsilon_{H}}{dq}=\left\langle \psi_{H}\left|\frac{d\hat{H}}{df_{H}}\right|\psi_{H}\right\rangle .\label{eq:curvature_Hellman-Feynman}
\end{equation}
The first equality is a restatement of Eq.\ (\ref{eq:Curvature Defnition}),
combined with the fact that the removed (added) charge is taken from
(inserted into) the highest occupied eigenstate $\psi_{H}$,~(see
footnote %
\footnote{Obviously the lowest-unoccupied eigenstate becomes the highest-occupied
one upon charge addition.%
}) while the second is due to the Hellmann-Feynman theorem~\citep{Hellman1937,Feynman1939}.
As the derivative in Eq.~\eqref{eq:curvature_Hellman-Feynman} is
applied only to the terms of the Hamiltonian $\hat{H}$ that are functionals
of the density, $n\left({\bf r}\right)$, we can write the curvature
as~\citep{Salzner2009} 
\begin{equation}
C=\int n_{H}\left(\mathbf{r}\right)\int\left[\frac{1}{|{\bf r}-\mathbf{r}'|}+f_{XC}\left({\bf r},\mathbf{r}'\right)\right]\frac{dn\left(\mathbf{r}'\right)}{df_{H}}d^{3}r'\mathrm{d}^{3}r,\label{eq:der}
\end{equation}
where $f_{XC}\left({\bf r},{\bf r^{\prime}}\right)=\delta^{2}E_{XC}[n]/\delta n(\mathbf{r})\delta n(\mathbf{r}')$
is the exchange-correlation kernel and $n_{i}\left(\mathbf{r}\right)\equiv\left|\psi_{i}\left({\bf r}\right)\right|^{2}$
is the density of the $i^{th}$ KS eigenstate. Using the fact that
the electron density is given by $n\left({\bf r}\right)=\sum_{i}f_{i}n_{i}\left(\mathbf{r}\right)$,
it follows that 
\begin{equation}
\frac{dn\left({\bf r}\right)}{df_{H}}=n_{H}\left({\bf r}\right)+n_{relax}\left(\mathbf{r}\right),
\end{equation}
where the first term on the right hand side is the density of the
highest occupied KS eigenstate and the second term, $n_{relax}\left(\mathbf{r}\right)\equiv\sum_{i}f_{i}\left(dn_{i}\left(\mathbf{r}\right)/df_{H}\right)$,
describes the eigenstate density relaxation upon charge removal/addition.
The curvature can therefore be expressed as:

\begin{equation}
\begin{aligned}C & =\iint\frac{n_{H}\left(\mathbf{r}\right)n_{H}\left(\mathbf{r}'\right)}{|{\bf r}-\mathbf{r}'|}d^{3}r'd^{3}r\\
 & \,\,+\iint\frac{n_{H}\left(\mathbf{r}\right)n_{relax}\left(\mathbf{r}'\right)}{\left|\mathbf{r}-\mathbf{r}'\right|}d^{3}r'd^{3}r+C_{XC}.
\end{aligned}
\label{eq:Curvature-Finite}
\end{equation}
The first term in Eq.~\eqref{eq:Curvature-Finite} is twice the electrostatic
interaction energy of $n_{H}\left(\mathbf{r}\right)$ with itself;
the second term is twice the electrostatic interaction energy between
$n_{H}\left(\mathbf{r}\right)$ and the relaxation density, $n_{relax}\left(\mathbf{r}\right)$;
the last term, 
\begin{equation}
C_{XC}=\iint n_{H}\left(\mathbf{r}\right)\left[n_{H}\left(\mathbf{r}'\right)+n_{relax}\left(\mathbf{r}'\right)\right]f_{XC}\left({\bf r},\mathbf{r}'\right)d^{3}r'd^{3}r,
\end{equation}
is the contribution of the exchange-correlation kernel to the curvature.
As discussed in the introduction, for the exact exchange-correlation
functional the curvature is identically zero and therefore the two
electrostatic (Hartree) terms must be canceled out by the exchange-correlation
kernel term.

In the LDA, the approximate exchange-correlation kernel is of the
form:~\cite{Petersilka1996} $f_{XC}^{LDA}\left(\mathbf{r},\mathbf{r}'\right)=\delta\left(\mathbf{r}-\mathbf{r}'\right)\tilde{f}_{XC}^{LDA}\left(n\left({\bf r}\right)\right)$.
The expression for the exchange-correlation contribution to the curvature
then simplifies to

\begin{equation}
C_{XC}^{LDA}=\int n_{H}\left(\mathbf{r}\right)\left[n_{H}\left(\mathbf{r}\right)+n_{relax}\left(\mathbf{r}\right)\right]\tilde{f}_{XC}^{LDA}\left(n\left({\bf r}\right)\right)d^{3}r,\label{eq:LDA-XC-Curvature}
\end{equation}
which does not generally cancel the Hartree terms in Eq.~\eqref{eq:Curvature-Finite}.
This is consistent with the above-mentioned deviations from piecewise-linearity
found in LDA calculations of small molecules.

\subsection{\label{sub:FinSys-EC-3D}Energy curvature in large finite three-dimensional
systems}

To gain insight into the behavior of curvature as a function of system
size, we first consider an electron gas consisting of $N_{e}$ electrons
distributed uniformly in a finite volume $\Omega$ with periodic boundary
conditions. For such a system, $n_{i}\left(\mathbf{r}\right)=\frac{1}{\Omega}$
and there is no eigenstate relaxation i.e. $n_{relax}(\mathbf{r})=0$.
Therefore the general curvature expression of Eq.~\eqref{eq:Curvature-Finite}
includes only the electrostatic self-interaction and XC terms and,
using LDA, can be simplified to

\begin{equation}
\begin{aligned}\bar{C} & =\frac{1}{\Omega^{2}}\iint_{\Omega}\frac{1}{|{\bf r}-\mathbf{r}'|}d^{3}r'd^{3}r+\frac{\tilde{f}_{XC}^{LDA}\left(n\right)}{\Omega}\\
 & =\frac{\bar{D}}{\Omega^{1/3}}+\frac{\tilde{f}_{xc}^{LDA}\left(n\right)}{\Omega},
\end{aligned}
\label{eq:HEG-Curvature}
\end{equation}
where we have used the fact that for a given uniform density, $n=N_{e}/\Omega$,
the LDA XC kernel is constant. Note that the bar over $C$ and $D$
is used to denote quantities relating to a uniform electron density.
The first term in Eq.\ \eqref{eq:HEG-Curvature} is twice the electrostatic
self-interaction energy of a unit charge, which is characterized by
a volume-independent shape factor $\bar{D}=\frac{1}{\Omega^{5/3}}\iint_{\Omega}\frac{1}{|{\bf r}-{\bf r^{\prime}}|}\mathrm{d}^{3}r'd^{3}r$.
Analytical integration yields $\bar{D}=\frac{6}{5}(\frac{4\pi}{3})^{1/3}E_{H}a_{0}\approx$52.5
$eVa_{0}$ for a sphere, where $E_{H}$ and $a_{0}$ are the atomic
Hartree and Bohr units for energy and length, respectively. For a
cube and a parallelepiped of the shape of a diamond primitive cell,
numerical integration yields $\bar{D}\approx$51.2 $eVa_{0}$ and
49.0 $eVa_{0}$, respectively, with the former value in agreement
with electrostatic energy calculations reported in Ref.~\citenum{finocchiaro1998}.

Clearly, the curvature of this uniform-electron-gas based example
decays to zero as the system size increases. Specifically, in the
limit of an infinitely large uniform electron gas limit, where LDA
is an exact result, the exact DFT condition of zero curvature is indeed
obeyed. However, for a uniform electron gas confined to a finite volume,
the LDA is not exact and therefore non-zero curvature is to be expected.

In Eq.~\eqref{eq:HEG-Curvature}, the curvature for large systems
is dominated by the $\bar{D}\Omega^{-1/3}$ term, which arises from
the electrostatic self-interaction of the highest occupied eigenstate.
It stands to reason that such a term, with a general prefactor $D$,
can be expected not just for this idealized system, but also for realistic
large but finite systems for which LDA is a reasonable approximation. 

To test this hypothesis, we focused on the elemental group IV solids
- diamond, silicon, and germanium - for which LDA is well-proven to
be a good approximation for ground-state properties.~\cite{Godby1987,Chelikowsky1992}
For each solid, we constructed a set of increasingly large nanocrystals
in two stages. First, we replicated the primitive unit cell of the
bulk crystal an equal number of times in each of the lattice vector
directions, using the experimental lattice constant, thereby creating
a finite but periodic supercell. Second, we removed unbound atoms
and passivated any remaining dangling bonds with hydrogen atoms. In
this way, hydrogen-passivated NCs containing up to 325 Si, C, or Ge
atoms, as well as a passivation layer containing up to 300 H atoms,
were formed. For each of the NCs constructed this way, we calculated
the LDA energy curvature for both charge removal and charge addition.
All calculations were performed using NWCHEM~\citep{Valiev2010}
with the cc-PVDZ basis set for the smaller NCs and the STO-3G basis
set for the larger NCs. The curvature was estimated by a finite difference
approximation to Eq.~\eqref{eq:Curvature Defnition}, $C=\Delta\varepsilon_{H}/\Delta f_{H}$,
where we calculated $\varepsilon_{H}$ for the neutral system and
for systems where an incremental small fractional charge was removed
from, or added to, the entire system.

\begin{figure}[h]
\begin{centering}
\includegraphics[width=8cm]{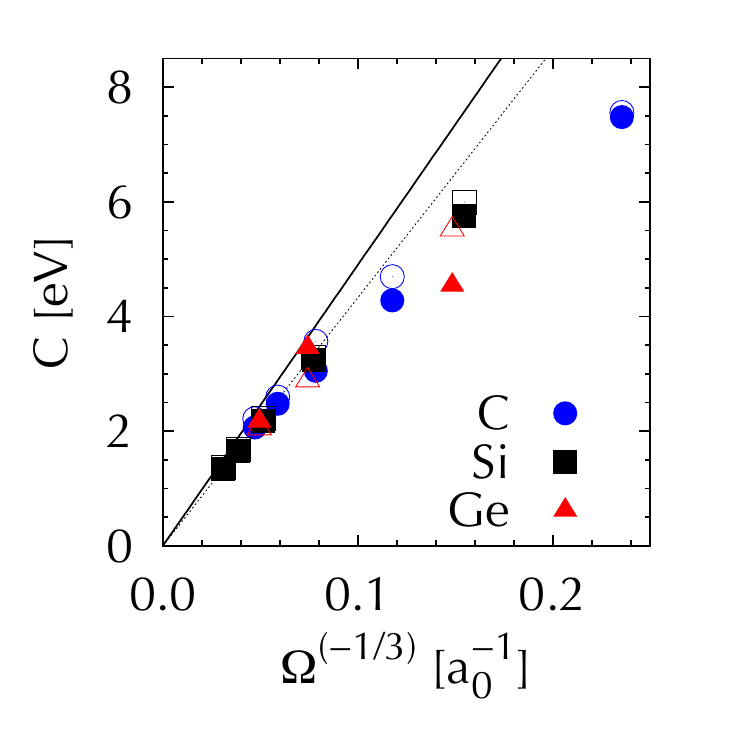} 
\par\end{centering}

\protect\caption{\label{fig:Curvatures-for-clusters} Curvature, $C$, obtained within
the local density approximation for electron removal (solid symbols)
or addition (hollow symbols) for diamond (blue circles), silicon (black
squares), and germanium (red triangles) nanocrystals, as a function
of $\Omega^{-\frac{1}{3}}$, where $\Omega$ is the nanocrystal volume.
The dotted line represents a least-squares fit to the asymptotic dependence.
The solid line represents the asymptotic dependence expected from
Eq.\ \eqref{eq:HEG-Curvature} for a uniform electron gas of the
same size and shape as the nanocrystals.}
\end{figure}

The resulting curvature for each of the systems studied is shown in
Fig.~\ref{fig:Curvatures-for-clusters}, as a function of $\Omega^{-\frac{1}{3}}$.
Clearly, in the limit of large system volume, $\Omega$, all three
compounds exhibit the limiting form expected, i.e., a curvature given
by $C=D\Omega^{-1/3}$, for both electron removal and addition. Furthermore,
by fitting our results for NCs with edges larger than $14a_{0}$ to
the expected dependence, we obtained $D\approx43.5eVa_{0}$ for all
three materials. This ``universal value'' is reasonable in light
of the fact that the highest occupied eigenstate for all three materials
has a similar spatial distribution, making the Hartree self-interaction
contribution similar. Moreover, it deviates from the ideal uniform-electron-gas
parallelepiped by only $\sim$20\%, a difference that can be attributed
to the non-uniform structure of the highest occupied eigenstate obtained
within LDA (see Eq.\ \eqref{eq:Curvature-Finite} above). For smaller
nanocrystals, the term scaling as $\Omega^{-1}$ is non-negligible
and therefore the curvature departs from the ideal $\Omega^{-1/3}$
behavior, as observed in Fig.~\ref{fig:Curvatures-for-clusters}.
Therefore, we conclude that the curvature expression given by the
right-hand side of Eq.\ \eqref{eq:HEG-Curvature}, derived for the
uniform electron gas, is indeed applicable also for realistic systems
possessing delocalized electronic states and that the self-repulsion
term dominates the curvature as the system grows.

Interestingly, further support for the limiting $D\Omega^{-1/3}$
dependence of the curvature is obtained from the results of past LDA-based
studies of the quantum size effect in spherical silicon \citep{Ogut1997}
and germanium \citep{Melnikov2004} nanocrystals. In these studies,
the difference between the fundamental gap, computed from total energy
differences of the anionic, neutral, and cationic system, was compared
to the KS eigenvalue gap. The difference was observed~\cite{Ogut1997,Godby1998,Melnikov2004}
to scale as $\sim\Omega^{-1/3}$. This observation is easily explained
within our theory as a direct consequence of the non-zero curvature:~\citep{Stein2012}
Eq.~\eqref{eq:Gaps_relations} shows that for (semi-)local functionals
(without an explicit derivative discontinuity), the difference between
the fundamental and the KS eigenvalue gap is in fact equal to the
average curvature for electron addition and removal and must exhibit
the same trends as a consequence. This conclusion is further supported
by the value of $D=39.5\, eVa_{0}$ and $D=41.1\, eVa_{0}$, deduced
for the spherical silicon and germanium NCs, respectively, from the
data of Ref.~\citenum{Ogut1997} and Ref.~\citenum{Melnikov2004}.
These values are indeed very close to the value of $D=43.5\, eVa_{0}$
which we obtained above from explicit curvature calculations for the
diamond-structured NCs. Note that the change in shape does not cause
a significant difference in the value of $D$, consistent with our
uniform electron gas calculations.

\subsection{\label{sub:EC-1D}Energy curvature in large finite one-dimensional
systems}

The above-demonstrated dominance of the electrostatic term in the
size-dependence of the curvature suggests that it must be strongly
influenced by dimensionality. To test this, we again consider twice
the Hartree energy as given in Eq.\ \eqref{eq:Curvature-Finite},
evaluated for a unit-charge uniform electron gas, confined to a cylinder
of length $L$ and radius $d$ such that $L\gg d$, as an approximation
for the curvature of a long but finite one-dimensional system. This
energy can be computed analytically~\citep{ciftja2012} to obtain:
\begin{equation}
C\approx\frac{2}{L}\ln\left(2e^{-3/4}\frac{L}{d}\right),\quad L\gg d.\label{eq:Cbar-lin}
\end{equation}
This indicates that, as in the three-dimensional case, the curvature
vanishes as the system grows arbitrarily long - an observation also
consistent with the results of Mori-S\textipa{á}nchez \emph{et al}.
for hydrogen chains \cite{Mori-Sanchez2008a}. However, the curvature
does not decay as $L^{-1}$, as perhaps could be naively expected,
but rather as $L^{-1}\ln\left(\frac{2e^{-3/4}L}{d}\right)$. The relaxation
and exchange-correlation terms are expected to scale as $L^{-1}$.
However they do not significantly affect the curvature when $L>50a_{0}$,
as for very large $L$ the logarithmic term dominates the $L^{-1}$
term\@.

\begin{figure}
\includegraphics{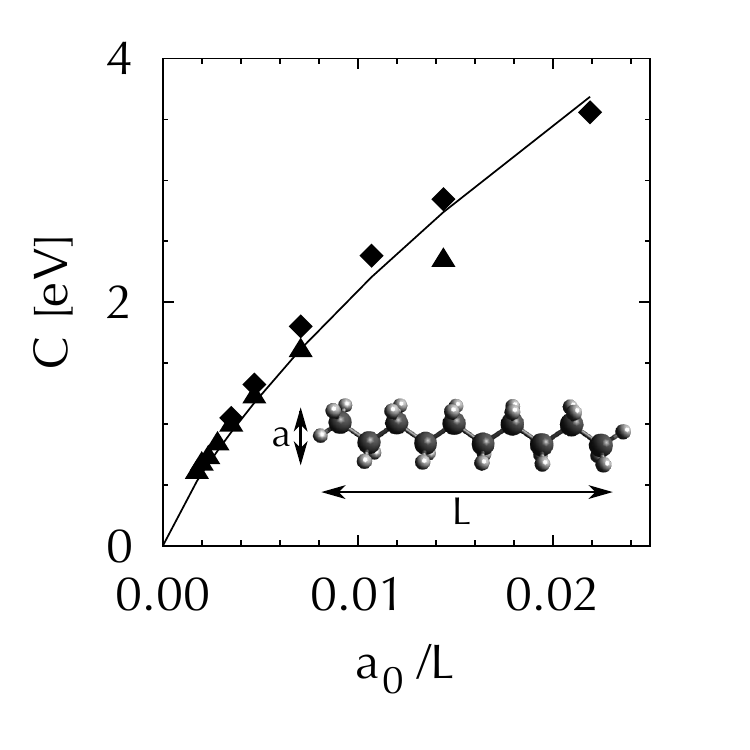}

\protect\caption{\label{fig:Curvature-c-chains} Curvature, $C$, obtained within the
local density approximation for electron removal from alkane chains,
as a function of $a_{0}/L$, where $L$ is the chain length. The line
represents the asymptotic dependence expected from Eq.\ \eqref{eq:Cbar-lin}
for a unit-charge uniform electron gas of the same length as the chain
and a radius of $d=2a_{0}$. Triangles and diamonds represent data
obtained using the cc-PVDZ and STO-3G basis sets, respectively. Inset:
the hexane molecule as an example of an alkane chain, with $d$ and
$L$ shown explicitly.}
\end{figure}

To test whether this prediction carries over to realistic one-dimensional
systems, we considered alkane chains of increasing length, $L$, whose
width $d$ is fixed by definition (see inset of Fig.~\ref{fig:Curvature-c-chains}).
These alkane chains provide a useful model of a quasi-one-dimensional
system that is well-described by LDA.~\citep{Segev2006} We investigated
chains containing up to 240 C atoms and again used NWCHEM~\citep{Valiev2010}
with the cc-PVDZ and the STO-3G basis sets. The computed curvature
for electron removal is shown in Fig.~\ref{fig:Curvature-c-chains},
as a function of $a_{0}/L$, and compared with the prediction of Eq.\ \eqref{eq:Cbar-lin}.
Clearly, for large $L$ the curvature is once again very well approximated
by the electrostatic self-interaction of a uniformly smeared unit
charge.

\section{Energy curvature in periodic systems}

\subsection{\label{sub:The-unit-cell-curvature}General Considerations}

In solid-state physics, it is common practice to employ periodic boundary
conditions for the description of crystalline solids.~\cite{Ashcroft2005}
To understand the bulk limit of curvature calculations in such a scenario,
we consider a reference cell of total volume $\Omega_{RC}$, containing
$N$ repeating unit cells (with unit cell volume $\Omega_{UC}$),
using Born - von Karman periodic boundary conditions.~\cite{Ashcroft2005}
In other words, the reference cell is treated as a \emph{finite but
topologically periodic} system.~\citep{kraisler2014fundamental}
In such a system, the infinite bulk limit is approached with increasing
size of the reference cell.(see footnote %
\footnote{Alternatively, the bulk limit can be approached using the concept
of $k$-point sampling of the unit cell.~\cite{Pickett1989} For
now, we do not pursue this alternative, but we discuss it extensively
below%
})

To compute the curvature, we remove (or add) an electronic charge
$Q_{RC}$ from (or to) the reference cell, denoted below by a ``hole''
(or ``elec'') superscript, where appropriate. We focus mostly on
electron removal for simplicity. Owing to the periodic boundary conditions,
electron removal from the reference cell implies removal of the same
charge from each of its periodic images. As there are an infinite
number of repeated reference cells, the removed electronic charge
is effectively infinite, leading to divergences in the Coulomb potential.
Therefore, a uniform compensating positive charge, of density $Q_{RC}/\Omega_{RC}$,
is introduced to the reference cell. This keeps the infinite periodic
crystal neutral and avoids the divergent behavior.~\citep{Ihm1979, Sharma2008}

As before, the curvature $C_{RC}$, defined with respect to the reference
cell, is computed as the rate of change of the highest occupied KS-eigenvalue
with respect to removed charge, $C_{RC}=d\varepsilon_{H}/dQ_{RC}$.
By construction (and assuming no symmetry breaking), the hole formed
by charge removal exhibits a periodic structure commensurate with
the repeating unit-cell and therefore $Q_{RC}=NQ_{UC}$, where $Q_{UC}$
is the charge removed from each of the $N$ unit cells that comprise
the reference cell. In the limit of large $N$, $\varepsilon_{H}$
is expected to become independent of the size of the reference cell,
i.e., of $N$. Therefore 
\begin{equation}
C_{RC}=\frac{d\varepsilon_{H}}{dQ_{RC}}=\frac{1}{N}\frac{d\varepsilon_{H}}{dQ_{UC}}\equiv\frac{C_{UC}}{N},\label{eq:Curvature-Infinite-Periodic}
\end{equation}
where $C_{UC}\equiv d\varepsilon_{H}/dQ_{UC}$ is the ``unit-cell
curvature'', which in the limit of large $N$ is independent of the
reference cell size (see footnote %
\footnote{The same dependence on $N$ can be obtained by considering the curvature
directly as the second derivative of the energy, i.e., $C_{RC}=d^{2}E_{RC}/dQ_{RC}^{2}$,
because $E_{RC}$ and $Q_{RC}$ are both extensive quantities and
therefore proportional to $N$.%
}).

Clearly, the curvature $C_{RC}$ for the infinite crystal does depend
on the reference cell size. As the reference cell grows ($N\to\infty$,
$\Omega_{RC}\to\infty$), we find $C_{RC}\rightarrow0$ for any underlying
functional. This result should be contrasted with the exact DFT condition
of piecewise linearity, where the curvature given by Eq.~\eqref{eq:Curvature-Infinite-Periodic}
should be strictly zero for \emph{any} reference cell size and not
just in the infinite cell limit. In other words, as for the NCs, in
the infinite system limit piecewise-linearity is obtained irrespective
of the underlying XC functional and therefore does not provide useful
information for functional construction or evaluation. However, in
the exact theory we also expect $C_{UC}=0$. Therefore, a non-vanishing
unit-cell curvature, $C_{UC}$, represents a measure of the spurious
XC functional behavior even in periodic infinite solids and may prove
useful in future analysis.

\subsection{\label{sec:LDA-of-Periodic} LDA calculations of topologically periodic
reference cells}

To examine the considerations and conclusions of the previous section,
we performed LDA calculations for increasingly large periodic reference
cells of selected semiconductors and insulators, using the LDA-optimized
lattice vectors of a neutral unit cell (see footnote %
\footnote{We performed non-spin-polarized calculations using norm-conserving
pseudopotentials within the Quantum-ESPRESSO \citep{QE-2009} and
ABINIT \citep{gonze2002first,gonze2005brief} packages, which use
a planewave basis with periodic boundary conditions. All results were
converged for plane-wave kinetic energy cut-off.%
}).

As mentioned above, the reference cell is considered to be finite
but topologically periodic. Therefore, all calculations are carried
out using only the single $k$-point (at $\Gamma$). This makes curvature
calculations straightforward both conceptually and practically, as
charge is removed from the highest occupied KS eigenstate as in the
finite-system calculations above. The energy derivatives needed for
the evaluation of the curvature (Eq.~\eqref{eq:Curvature-Infinite-Periodic})
were calculated using finite differences of the highest occupied energy
eigenvalue, $\varepsilon_{H}$, for the neutral and incrementally
charged reference cell.

\begin{figure}[h]
\includegraphics[width=8cm]{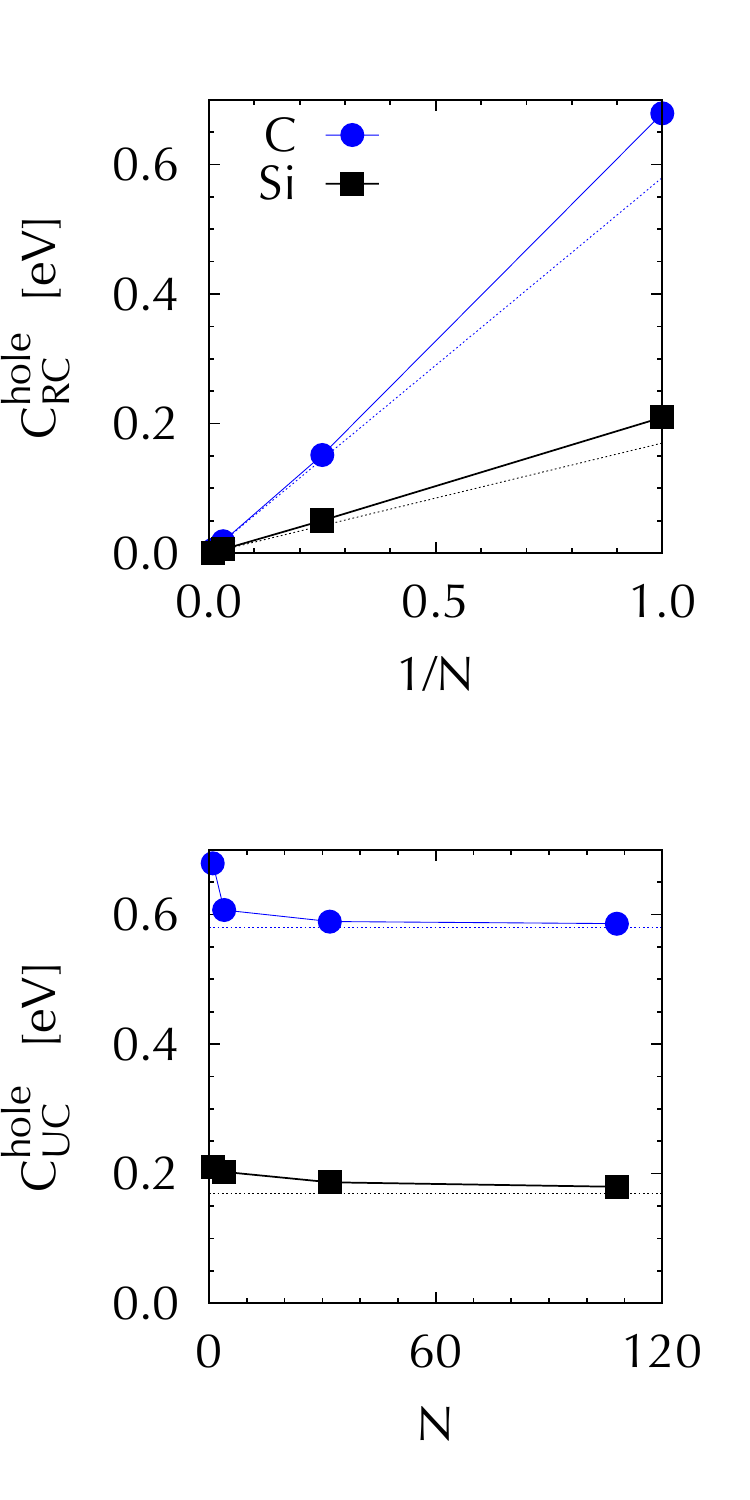} \protect\caption{\label{fig:Curvature-CN-for-Si} Computed charge removal curvature
for silicon (black squares) and diamond (blue circles) crystals: (Top)
Reference cell curvature, $C_{RC}^{hole}$, as a function of $N^{-1}$.
(Bottom) unit cell curvature, $C_{UC}^{hole}$, as a function of $N$,
where $N$ is the number of primitive unit-cells in the reference
cell. Solid Lines are a guide for the eye. Dotted lines represent
the asymptotic dependence on $N$ (Top), or the converged values of
$C_{UC}$ (Bottom), obtained through Brillouin-zone sampling described
in Sec. \ref{sub:Brillouin-zone-sampling}.}
\end{figure}

The results of such calculations for increasingly large reference
cells of diamond and silicon are summarized in Fig.~\ref{fig:Curvature-CN-for-Si}.~(see
footnote %
\footnote{LDA erroneously predicts bulk germanium to be semi-metallic (see,
e.g., J. Heyd, J. E. Peralta, G. E. Scuseria, and R. L. Martin, J.
Chem. Phys. 123, 174101 (2005)). Therefore, germanium is omitted from
Fig.~\ref{fig:Curvature-CN-for-Si}%
}) As shown in the top panel of Fig.\ \ref{fig:Curvature-CN-for-Si},
the reference cell curvature indeed decreases monotonically and vanishes
in the large $N$ limit, in agreement with the above theoretical considerations.
At the same time, the bottom panel of Fig.\ \ref{fig:Curvature-CN-for-Si}
shows that the unit cell curvature is not zero and for large $N$
approaches a constant, material-dependent value, such that Eq.\ \eqref{eq:Curvature-Infinite-Periodic}
is obeyed.

\subsection{\label{sec:The-finite-infinite-paradox} Finite versus periodic cell:
A seeming paradox and its resolution}

In the limit of an arbitrarily large system, one would expect surface
effects to be negligible and so, naively, that the limiting behavior
of large periodic and non-periodic systems to be the same. However,
we already showed both analytically and numerically that in fact the
limiting behavior is not the same. For the finite system, the curvature
asymptotically scales as $\Omega^{-1/3}$, where $\Omega$ is the
volume of the finite, non-periodic system, whereas for the topologically
periodic system the curvature asymptotically scales as $\Omega_{RC}^{-1}$,
where $\Omega_{RC}$ is the reference cell volume.

This apparent paradox can be reconciled by recalling that in a periodic
system, electron addition/removal must be accompanied by the addition
of a compensating, uniformly distributed background charge of opposite
sign, so as to avoid divergence of the Coulomb potential and energy.~\citep{Ihm1979}
For a non-periodic system, however, no compensating charge is necessary.
This background charge strongly affects curvature considerations.\cite{Sharma2008}
To understand why, consider that if surface effects are neglected
then Eq.\ \eqref{eq:Curvature-Finite}, developed above for non-periodic
systems, can be applied to the reference cell of a periodic system.
However, while $n_{relax}$ integrates to zero in the reference cell,
$n_{H}$ integrates to 1. Therefore, $n_{H}$ must be replaced by
a background-neutralized density, $\rho_{H}\left(\mathbf{r}\right)\equiv n_{H}\left(\mathbf{r}\right)-\frac{1}{\Omega_{RC}}$,
before it can be inserted in Eq.\ \eqref{eq:Curvature-Finite}. Therefore,
Eq.\ \eqref{eq:Curvature-Finite} yields the following expression
for the curvature in the periodic case, $C^{periodic}$: 
\begin{equation}
\begin{aligned}C^{periodic} & =\iint_{\Omega_{RC}}\frac{\rho_{H}\left(\mathbf{r}\right)\rho_{H}\left(\mathbf{r'}\right)}{|{\bf r}-\mathbf{r}'|}d^{3}r'd^{3}r\\
 & \,\,+\iint_{\Omega_{RC}}\frac{\rho_{H}\left(\mathbf{r}\right)n_{relax}\left(\mathbf{r'}\right)}{\left|\mathbf{r}-\mathbf{r}'\right|}d^{3}r'd^{3}r+C_{XC}.
\end{aligned}
\label{eq:RefCell-Curvature}
\end{equation}

With all densities being unit-cell periodic, we can define $\tilde{n}_{j}\left(\mathbf{G}\right)=\frac{1}{\Omega_{UC}}\int_{\Omega_{UC}}n_{j}\left(\mathbf{r}\right)e^{i\mathbf{G}\cdot\mathbf{r}}d^{3}r$
as the Fourier-component of $n_{j}\left(\mathbf{r}\right)$ corresponding
to the reciprocal unit-cell lattice vector, $\mathbf{G}$. For charge-neutral
systems, the $\mathbf{G}=\mathbf{0}$ component must be zero. By noting
that $\tilde{n}_{H}\left(\mathbf{G}\neq\mathbf{0}\right)=\tilde{\rho}_{H}\left(\mathbf{G\neq0}\right)$,
because the two densities differ only by a constant, we obtain: 
\begin{equation}
\begin{aligned}C^{periodic} & =4\pi\Omega_{RC}\sum_{\mathbf{G}\ne0}\frac{\tilde{n}_{H}\left(\mathbf{G}\right)\tilde{n}_{H}\left(\mathbf{G}\right)^{*}}{G^{2}}\\
 & \,\,+4\pi\Omega_{RC}\sum_{\mathbf{G}\ne0}\frac{\tilde{n}_{H}\left(\mathbf{G}\right)\tilde{n}_{relax}\left(\mathbf{G}\right)^{*}}{G^{2}}+C_{XC}.
\end{aligned}
\label{eq:RC-Curvature-G}
\end{equation}

The KS-eigenstate densities, $n_{j}\left(\mathbf{r}\right)$, are
normalized over the reference cell and therefore $n_{H}\left(\mathbf{r}\right)$
and $n_{relax}\left(\mathbf{r}\right)$, as well as their Fourier
components, must scale as $\Omega_{RC}^{-1}$. Because \textbf{$\mathbf{G}$}
depends only on the unit cell and is independent of $\Omega_{RC}$,
Eq.\ \eqref{eq:RC-Curvature-G} shows that $C^{periodic}$ scales
as $\Omega_{RC}^{-1}$. Thus, we obtain a curvature that scales with
inverse system volume, consistent with Eq.\ \eqref{eq:Curvature-Infinite-Periodic}
above.

One can also compare the terms in Eq.\ \eqref{eq:Curvature-Finite}
and Eq.~\eqref{eq:RefCell-Curvature}, obtaining the following expression
for the difference in their curvature: 
\begin{equation}
\begin{array}{l}
C^{finite}-C^{periodic}\\
\,\,\,\,\,\,\,\,\,\,=\iint_{\Omega_{RC}}\frac{\left(2n_{H}\left(\mathbf{r}\right)-\frac{1}{\Omega_{RC}}-n_{relax}\left(\mathbf{r}\right)\right)\frac{1}{\Omega_{RC}}}{|{\bf r}-\mathbf{r}'|}d^{3}r'd^{3}r.
\end{array}\label{eq:Finite-Periodic-C-Diff}
\end{equation}
Dimensional analysis reveals that the above curvature difference scales
as $\Omega_{RC}^{-1/3}$. 

As discussed in Section \ref{sec:EC-finite}, $\Omega_{RC}^{-1/3}$
scaling was also obtained for the non-periodic case from the self-interaction
energy of the highest-occupied eigenstate. Furthermore, because the
background charge must systematically cancel the divergence in the
electronic electrostatic energy, the prefactor of the $\Omega_{RC}^{-1/3}$
dependence in the above equation must be equal and opposite to that
deduced from Eq.\ \eqref{eq:Curvature-Finite}. Therefore, overall
the $\Omega_{RC}^{-1/3}$ scaling must vanish in $C^{periodic}$ and
only the $\Omega_{RC}^{-1}$ scaling remains.

To summarize, the scaling behavior of a non-periodic and a periodic
system really is different, but this is not owing to topology \emph{per
se}, but rather stems from the effects of the uniform background charge,
used in periodic calculations only. Before concluding this issue,
however, two more comments are in order. First, for finite systems
described with (semi-)local functionals, the scaling is self-interaction
dominated and therefore positive (see, e.g., Refs.\ \citenum{Mori-Sanchez2006,
Cohen2008a, Cohen2008c, Stein2012}). Upon elimination of this effect
by the compensating background, curvature can be either positive or
negative (which we show below to be the case). This is somewhat reminiscent
of the behavior of the exact-exchange functional (see, e.g., Ref.\ \citenum{Mori-Sanchez2006}),
where self-interaction is eliminated and the curvature is typically
mildly negative. Second, for periodic systems we assumed throughout
that the removed/added charge is delocalized throughout the reference
cell. If this is not the case, e.g., if a molecule or a localized
defect is computed within a large supercell, scaling arguments no
longer apply and the results will resemble those of finite systems.
This explains, among other things, why a Hubbard-like $U$ term for
localized states in an otherwise periodic system is indeed useful,
as long as the correction is limited to the vicinity of the localized
site (see, e.g., Refs.\citenum{Solovyev1994,cococcioni2005linear}).

\subsection{\label{sub:Brillouin-zone-sampling}Brillouin zone sampling}

In Section \ref{sec:LDA-of-Periodic} we have considered the infinite
solid limit by constructing increasingly large topologically-periodic
reference cells. While pedagogically useful, this procedure is too
cumbersome and computationally expensive to be used for routine unit-cell
curvature calculations. In practice, the infinite-solid limit is much
easier to reach by using $k$-point sampling of the Brillouin zone
corresponding to a single periodic unit cell. This sampling relies
on Bloch's theorem, which allows the eigenfunctions of a periodic
Hamiltonian to have the same periodicity up to a phase factor.~\cite{Ashcroft2005}
One can then show that a single unit cell with uniform sampling of
$N$ $k$-points is completely equivalent, mathematically and physically,
to a reference cell comprised of $N$ unit cells within the single
$k$-point ($k=0$) treatment.~\citep{Pickett1989} Specifically,
for a semiconductor or insulator in the ground state the $n$ electrons
in the unit cell occupy the lowest eigenstates, with energies $\varepsilon_{i,k}$
(where $i$ is the band index and $k$ is the $k$-point index). This
is equivalent to a system with $n\times N$ electrons, i.e. a reference
cell comprised of $N$ unit cells. The infinite solid limit, then,
simply corresponds to an arbitrarily dense $k$-point sampling.

Obviously, practical calculations must involve a finite number of
$k$-points. This is of little consequence to ground-state calculations
of semiconductors and insulators, as results tend to converge quickly
with the number of $k$-points.~\citep{Pickett1989} However, it
raises a serious issue for electron removal/addition calculations.

Naively, one would think that the above-discussed determination of
curvature from $d\varepsilon_{H}/dq$ should be generalized to the
case of $k$-point sampling by considering $d\varepsilon_{F}/dq$,
where $\varepsilon_{F}$ is the Fermi level. This is because for a
ground-state, zero-temperature solid, $\varepsilon_{F}$ denotes the
energy of the highest occupied state by definition. However, in practice
one always removes/adds a finite amount of charge, $q$, rather than
a truly infinitesimal charge. Therefore, charge is generally removed
from all eigenstates with energy $\varepsilon_{i,k}$ greater than
$\varepsilon_{F}$, where the latter is determined by the charge conservation
condition 
\begin{equation}
nN-q=\int_{-\infty}^{\varepsilon_{F}}g\left(\varepsilon\right)d\varepsilon,\label{eq:Finding-Eps_F}
\end{equation}
where $g\left(\varepsilon\right)$ is the density of states (DOS).
Once charge is removed not only from the highest-energy state, but
rather from many states, the piecewise linearity condition no longer
applies. Therefore the entire theoretical edifice on which all previous
considerations were based breaks down. This difficulty persists even
if the second derivative of the total energy, rather than the first
derivative of the Fermi energy, is considered. One could, perhaps,
hope that extrapolation of $d\varepsilon_{F}/dq$ to $q\to0$, where
charge really is removed only from the highest occupied eigenstate,
would still lead to the correct result. Unfortunately, this is not
the case. For example, within the effective mass approximation it
is well-known that that $\varepsilon_{F}-\varepsilon_{H}\sim q^{2/3}$,
where $\varepsilon_{H}$ is the top of the valence band.~\citep{Kittel1966}
Clearly, then, $d\varepsilon_{F}/dq$ actually diverges for $q\to0$.

\begin{figure}[h]
\includegraphics[width=8cm]{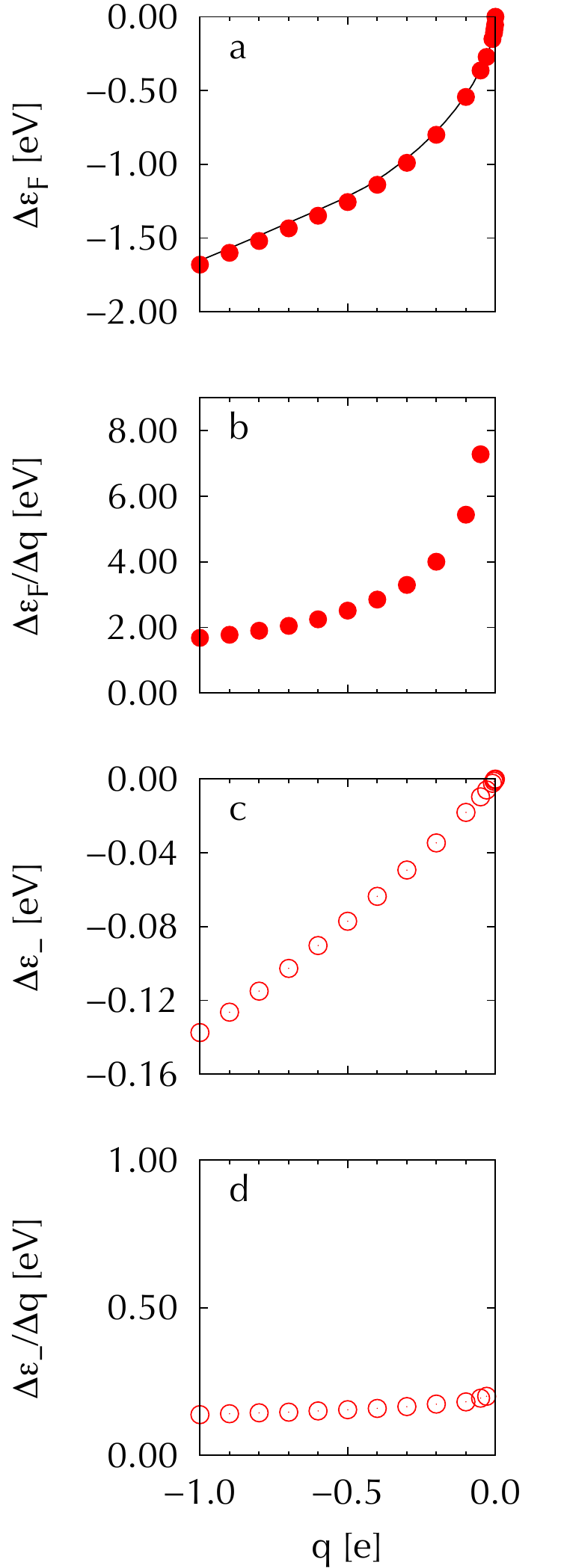}

\protect\caption{\label{fig:fermi_energy} Charge removal in a unit cell of silicon,
with $16\times16\times16$ $k$-point sampling. Panel a: change in
Fermi level position, $\Delta\varepsilon_{F}$, as a function of the
removed charge, $\Delta q$. Solid line: change in $\varepsilon_{F}$
expected from the uncharged density of states curve. Panel b: Numerical
derivative of the data in panel a, $\Delta\varepsilon_{F}/\Delta q$,
as a function of $\Delta q$. Panel c: change in position of valence
band maximum, $\Delta\varepsilon_{-}$, as a function of $\Delta q$.
Panel d: Numerical derivative of the data in panel c, $\Delta\varepsilon_{-}/\Delta q$,
as a function of $\Delta q$. }
\end{figure}

The above considerations are illustrated numerically in Fig.~\ref{fig:fermi_energy},
where the dependence of $\varepsilon_{F}$ on $q$ (a) and its derivative
(b) were computed for a primitive unit cell of silicon with a $16\times16\times16$
$k$-point sampling scheme. Clearly, and as expected from Eq.\ \eqref{eq:Finding-Eps_F},
the Fermi energy follows closely the integrated density of states
of the uncharged system (shown as a solid line). Furthermore, for
small $q$ it indeed follows a $q^{2/3}$ law and its derivative diverges.

Fortunately, an equally simple, yet accurate, procedure is to consider
instead the valence band maximum (or the conduction band minimum for
charge addition), which we denote here as $\varepsilon_{-}$. For
$q\to0$ it too must tend to the correct limit as charge is removed
only from the highest occupied state. For finite $q$ it is, of course,
incorrect, but as it does not incorporate DOS effects its derivative
is not expected to diverge. This is illustrated numerically in Fig.~\ref{fig:fermi_energy}
as well, for the same silicon example, where both the weaker dependence
of $\varepsilon_{-}$ on $q$ (c - note scale) and the convergence
of its derivative for small $q$ (d) is apparent.

\begin{figure}[h]
\centering{}\includegraphics[width=8cm]{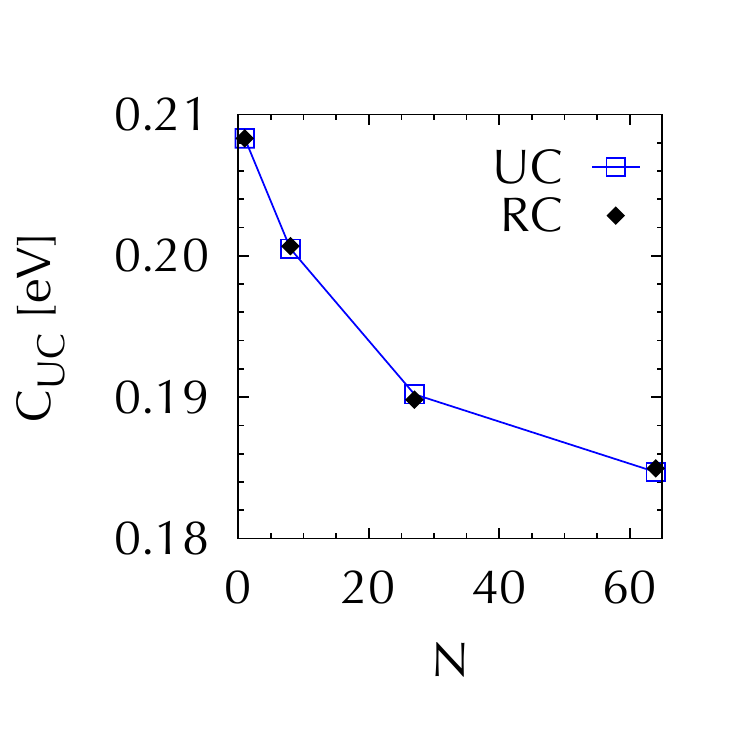} \protect\caption{\label{fig:convergence} Unit cell curvature, $C_{UC}$, obtained
from single $k$-point calculations of a reference cell containing
$N$ unit cells (filled diamonds) and from $N$ $k$-point calculations
of a single unit cell (hollow squares), as a function of $N$. Lines
joining the results of the $N$ $k$-point calculations are a guide
to the eye.}
\end{figure}

In the calculations of Fig.\ \ref{fig:fermi_energy}, the removal
of charge $q$ from a unit cell, sampled by $N$ $k$-points, is in
fact equivalent to the removal of the same charge from a reference
cell whose volume is $N$ times larger. Using Eq.\ \eqref{eq:Curvature-Infinite-Periodic},
this means that the limiting value of $\Delta\varepsilon_{-}/\Delta q$
is directly comparable to the non-vanishing unit-cell curvature, $C_{UC}$,
rather than to the vanishing reference cell curvature, $C_{RC}$.
This is directly verified in Fig.\ \ref{fig:convergence}, which
compares, for silicon, unit-cell curvature values, $C_{UC}$, obtained
from increasingly large single $k$-point reference cells (as in Fig.\ \ref{fig:Curvature-CN-for-Si})
with those obtained from increasingly dense $k$-point sampling of
a unit cell. Clearly, the results are indeed equivalent.

Finally, with the above scheme, we efficiently calculate unit-cell
curvatures for charge removal and addition in a variety of semiconductors
and insulators, obtained in the limit of sufficiently dense $k$-point
sampling. The results are summarized in Table \ref{tab:Energy-curvature}.

From the results it is clear that for all systems considered the unit
cell curvature in LDA is a non zero, material-dependent property.
Furthermore, once convergence has been reached it is independent of
the density of the $k$-point sampling. This is to be contrasted with
the reference cell curvature discussed earlier, which was not only
dependent on the reference cell size, but went to zero in the infinite
limit for all functionals. As noted in the preceding section, $C_{UC}$
can have both positive and negative values (illustrated by the results
in \ref{tab:Energy-curvature}), owing to the presence of the neutralizing
background.

\begin{table}[h]
\begin{tabular}{|l|l|c|c|}
\hline 
\multirow{2}{*}{} & \multirow{2}{*}{Crystal Structure } & \multicolumn{2}{c|}{$C_{UC}\,\left(eV\right)$}\tabularnewline
\cline{3-4} 
 &  & Charge removal  & Charge addition \tabularnewline
\hline 
AlAs  & Zinc-blende  & 0.26  & -0.65 \tabularnewline
AlN  & Zinc-blende  & 0.94  & -0.92\tabularnewline
AlP  & Zinc-blende  & 0.33  & -0.66\tabularnewline
AlSb  & Zinc-blende  & 0.16  & -0.57\tabularnewline
C  & Diamond  & 0.58  & -0.62\tabularnewline
GaP  & Zinc-blende  & 0.36  & -0.65\tabularnewline
MgO  & Rock-salt  & 1.6  & -0.73 \tabularnewline
Si  & Diamond  & 0.17  & -0.54 \tabularnewline
SiC  & Zinc-blende  & 0.59  & -0.64\tabularnewline
\hline 
\end{tabular}\protect\caption{\label{tab:Energy-curvature}Energy curvature for the unit cell, $C_{UC}$,
for charge removal and addition, calculated for various solids.}
\end{table}

\section{Conclusions}

In this article, we have examined the solid-state limit of energy
curvature, i.e., of deviations from piecewise-linearity, focusing
on (semi-)local functionals. We considered two different limits: finite
systems, with volume $\Omega\to\infty$, as well as topologically
periodic systems with a reference cell (to which the periodic boundary
conditions are applied) of volume $\Omega_{RC}\to\infty$. We found
that in all cases piecewise-linearity - albeit possibly with the wrong
slope - is obtained in the solid-state limit, even from functionals
that grossly disobey it for a finite system. However, using both analytical
considerations and practical calculations of representative systems,
we found that this limit is reached in very different ways. Therefore,
while using the demand of zero curvature for functional construction
and evaluation is not, as such, useful in the solid-state limit, its
size-dependence does contain useful information.

For large finite systems, we found that curvature scales as $\Omega^{-1/3}$
for three-dimensional systems (e.g., nanocrystals) and as $\frac{2}{L}\ln\left(2e^{-3/4}\frac{L}{a}\right)$,
where $L$ is the length and $a$ is the width, for quasi-one-dimensional
systems (e.g., molecular chains). This scaling behavior was found
to be dominated by electrostatics and traced to the self-interaction
term of the highest occupied state.

For large reference cell periodic systems, we found that the curvature
$C_{RC}$ scales as $C_{RC}=C_{UC}\Omega_{UC}/\Omega_{RC}$, where
$C_{UC}$ and $\Omega_{UC}$ are the unit-cell curvature and volume
respectively. $C_{UC}$ (for an approximate functional) is a non-vanishing
material-dependent quantity that is independent of the reference cell,
and therefore may serve as a new useful measure of functional error
in periodic solids, similar to that of the deviation from piecewise
linearity used in finite systems. Furthermore, we have been able to
calculate this curvature in two ways: either directly from the definition
by using increasingly large periodic cells or, more usefully, by considering
changes in the band edge position with increasingly dense $k$-point
sampling. Last but not least, we rationalized the difference between
the periodic and non-periodic case as resulting from the automatic
elimination of the spurious self-interaction via the addition of a
compensating background charge in periodic system.

We believe that these results should prove useful for further development,
evaluation, and application of novel exchange-correlation functionals
suitable for the solid-state.

\section{Acknowledgments}

We thank Eli Kraisler, Sivan Refaely-Abramson (Weizmann Institute),
and Stephan Kümmel (Universität Bayreuth) for useful discussions.
Work at the Weizmann Institute was supported by the European Research
Council. Work at the Hebrew University of Jerusalem was supported
by the Israel Science Foundation Grant No. 1219-12. Some of the computations
(VV) were performed at the Leibniz Supercomputing Centre of the Bavarian
Academy of Sciences and the Humanities. 

\bibliographystyle{apsrev4-1}
\bibliography{RoiBaerLib}

\end{document}